\documentclass[reprint,
superscriptaddress,
amsmath,amssymb,aps,showkeys,showpacs,
twoside,final,secnumarabic,
nofootinbib,twocolumn]{revtex4-2}

\usepackage[paperwidth=205mm,paperheight=290mm,top=17mm,bottom=25mm,
inner=17mm,outer=17mm,
twoside]{geometry}

\usepackage{cmap} 
\usepackage[T1,T2A]{fontenc}
\usepackage[utf8]{inputenc}
\usepackage[russian,english]{babel}
\usepackage{color}
\usepackage{graphicx}
\usepackage{dcolumn}
\usepackage{bm} 
\usepackage[unicode=true,colorlinks=true,linkcolor=magenta, urlcolor=blue, citecolor = blue,breaklinks]{hyperref}
\usepackage{multirow}
\usepackage{url}
\usepackage{breakurl}
\DeclareGraphicsExtensions{.eps}

\newcount\issue
\newcount\Vol
\newcount\numb
\usepackage{fancyhdr} 
\def\Vol{\textbf{80}}
\def\numb{x}
\begin{document}

\title{JOURNAL SECTION OR CONFERENCE SECTION \\[20pt]
HII regions in NGC 628: the view of two catalogs} 

\def\addressa{Ural Federal University}
\def\addressb{Institute of Astronomy of the Russian Academy of Sciences}

\author{\firstname{K.I.}~\surname{Smirnova}}
\email[E-mail: ]{Arashu@rambler.ru }
\affiliation{\addressa}
 \author{\firstname{D.S.}~\surname{Wiebe}}
\affiliation{\addressb}

\received{xx.xx.2025}
\revised{xx.xx.2025}
\accepted{xx.xx.2025}

\begin{abstract}
The study is devoted to comparing the parameters of the interstellar medium of HII regions in the Kongiu and Groves catalogs  for the galaxy NGC 628. The article analyzes the characteristics of star-forming regions, including a comparison of radiation fluxes in the ranges of 7.7 $\mu$m and 21 $\mu$m and in the H$\alpha$, H$\beta$, OIII and CO lines, calculating the kinematic parameters (FWHM) for the lines, and analyzing the spatial distribution of regions for both catalogs.

The results of the study showed that the regions from the Groves catalog demonstrate higher line widths compared to the Kongiu catalog. Signs of possible misidentified classification of some regions from the Groves catalog were revealed: there is a possibility that some of them are not HII regions, but shock ionization regions.
 
\end{abstract}

\par
\keywords{Suggested keywords   \\[5pt]}

\maketitle
\thispagestyle{fancy}

\section{Introduction}\label{intro}

The nature and driving forces behind a large-scale star formation in disk galaxies are still a subject of debate. One way to disentangle processes shaping gas and stellar components of star-forming regions is to relate observational tracers of various dust species and of ionized, atomic, and molecular gas. Such studies have been performed by numerous groups both on a pixel-by-pixel basis and on the molecular cloud scales \cite{1,2} for different samples of nearby disk galaxies.

NGC 628 is a star-forming spiral galaxy that is well-suited for such studies. It is oriented almost face-on, which reduces confusion effects. An ample amount of both low-resolution and high-resolution data are available for this galaxy, including such surveys as  SINGS~\cite{3} (3.6, 4.5, 5.8, 8.0 and 24 $\mu$m, Spitzer Space Telescope), HERACLES~\cite{4} (CO (2–1) line, IRAM),  PHANGS-ALMA survey~\cite{5,6} (CO (2–1) line, ALMA), PHANGS-JWST~\cite{7} (2, 3, 3.35, 3.6, 7.7, 10, 11.3, 21 $\mu$m), PHANGS-MUSE survey~\cite{8} (H$\alpha$ and H$\beta$ line and OIII line). It has also been a subject of our own previous studies \cite{9,10}. In these studies, we relied on the visual detection method as a tool to identify star-forming complexes in the studied galaxies. This approach was effective when working with low-resolution data. However, with an advent of high-resolution data, there is a need for more sophisticated methods. Such methods have been invoked recently to extract locations of HII regions in a number of disk galaxies, including NGC~628, from the MUSE survey data~\cite{8} and resulted in two catalogs, published by Congiu et al.~\cite{11} and Groves et al.~\cite{12} (hereinafter the Congiu catalog and the Groves catalog, respectively). In this contribution, we compare these two catalogs and make conclusions on their similarity and differences.

\section{\label{sec:level1}Catalogs}

The catalogs by Congiu et al.~\cite{11} and Groves et al.~\cite{12}) are built on the same data, but rely on different methods to identify HII~regions. Surprisingly, the catalogs differ somewhat both in the number of detected regions and in their spatial locations. This situation prompted us to perform a comparative analysis of these catalogs using the galaxy NGC~628 as an example.

In the Congiu catalog \cite{11}, the CLUMPFIND \cite{13} package was used for selection. The regions were selected by their H$\alpha$, OIII $\lambda=$5007, and S II $\lambda=$6717, 6731 emission. The adopted selection criteria allowed distinguishing between regions of ionized gas of various nature, including shocks and planetary nebulae.

In the Groves catalog \cite{12}, the HII-phot program \cite{15} was used. This program is specially designed to extract HII regions based on H$\alpha$ maps. The peaks of H$\alpha$ emission are identified and become the centers of the regions from which the expansion of the regions begins. The regions are then expanded using iterative Gaussian smoothing. Boundaries are set where the emission becomes too weak to be classified as HII regions. The accent was placed only on HII regions that were selected by their H$\alpha$ emission.

In the Congiu catalog, 3502 ionized regions are identified in NGC~628, with 2798 regions classified as `true' HII regions. The Groves catalog contains information about 2855 HII regions.

\section{\label{sec:level1}RESULTS}

\subsection{\label{sec:citeref}Location of HII regions}

\begin{figure}[h]
\centering{\includegraphics[width=0.5\textwidth]{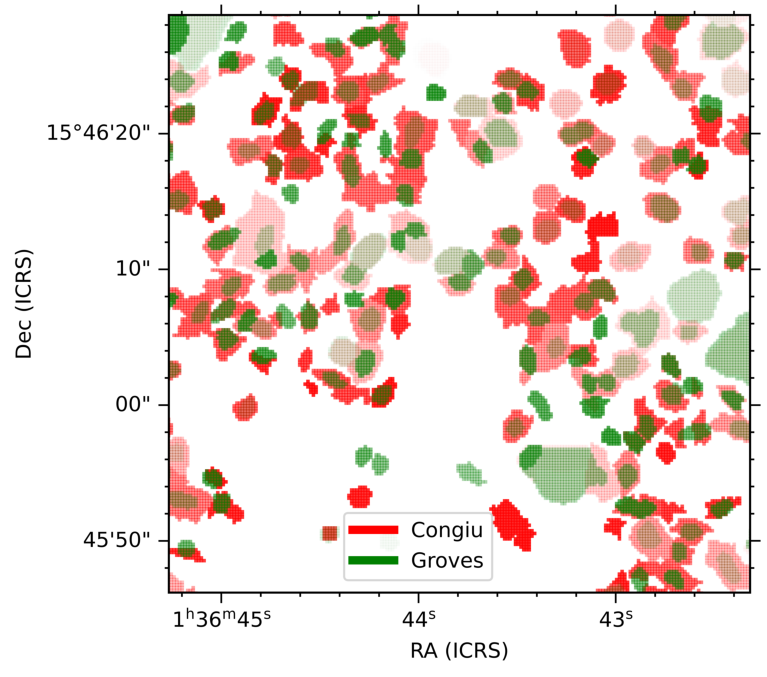}}
\caption{Locations of HII regions in NGC 628 according to the Congiu and Groves catalogs.}
\label{fig:NGC628}
\end{figure}

In Fig.~\ref{fig:NGC628} we show a fragment of the NGC~628 HII region map for both catalogs, with the Congiu regions shown in shades of red and the Groves regions shown in shades of green. It is seen on this plot that the catalogs are in general similar to each other, but with some notable differences.

First of all, the plot is more red than green, that is the Congiu regions are more spatially extended even when their locations are consistent with the locations of the Groves regions. One way to characterize this is to compare the average area of the regions both catalogs, which is 62 pixels in the Groves catalog and 90 pixels in the Congiu catalog. Only 91 regions from the Congiu catalog (3\% of its HII regions) are completely overlapped with a single region from the Groves catalog, while in the Groves catalog 1186 regions (42\%) are completely overlapped by the Congiu catalog regions.

There are also significant differences in the identification of HII regions. In Fig.~\ref{fig:NGC628} we see regions from different catalogs that overlap only partially or do not overlap at all. In other words, there are regions, which are only identified in one of the catalogs. The Congiu catalog contains 432 unique regions (15\% of their total number), which do not overlap with regions from the Groves catalog, while Groves has 180 such unique regions (6\% of the total). In both cases, unique regions are those with small surface brightness in H$\alpha$. This may signify that both algorithms produce more consistent results for brighter regions and become more ambiguous for fainter regions.

Analysis of the mosaic overlap shows that in the Congiu catalog 214 regions (7\%) are overlapped by several regions from the Groves catalog, while in the Groves catalog the number of mosaically overlapped regions is 417 (14\%).

\subsection{\label{sec:citeref}Fluxes and line profiles}

We compared fluxes at 7.7 $\mu$m and 21 $\mu$m, as well as fluxes in H$\alpha$ and CO lines for the regions in both catalogs. The purpose of the comparison was to identify differences in relations between the fluxes depending on the method of region selection. The analysis showed a good correlation between the fluxes in the near- and mid-IR ranges (7.7 $\mu$m and 21 $\mu$m) for the regions of both catalogs. All regions fall on a single trend line. A similarity was also observed when comparing the fluxes in the near- and mid-IR ranges with fluxes both in H$\alpha$ line and the CO line.

In addition to the analysis of the fluxes and surface brightnesses of the two catalogs, a study of the kinematic characteristics of the regions was conducted. Previously, we employed a $\delta V$ parameter, which characterizes the velocity scatter of the region, for such an assessment. The use of this parameter was warranted by large sizes of considered star-forming complexes and the possible presence of several kinematic components that cannot be correctly described by a single Gaussian.

In this paper, we study much more compact regions with a single kinematic component, which allows us to fit the line profiles with a single Gaussian. For each line under consideration, the line spread function was calculated using formula (3) from \cite{8}. After that, the instrumental profile was subtracted. Ultimately, we obtained the Gaussian parameters, $\sigma$ (converted into FWHM), the line center, and the background level. We calculated the FWHM for H$\alpha$, H$\beta$, and OIII lines. 

When comparing the surface brightnesses and FWHM of H$\beta$ and OIII lines, we notice a weak positive correlation between H$\beta$ surface brightness and H$\beta$ FWHM and a weak negative correlation between OIII surface brightness and OIII FWHM, nearly the same in both catalogs.

\begin{figure}[b]
\includegraphics[width=0.5\textwidth]{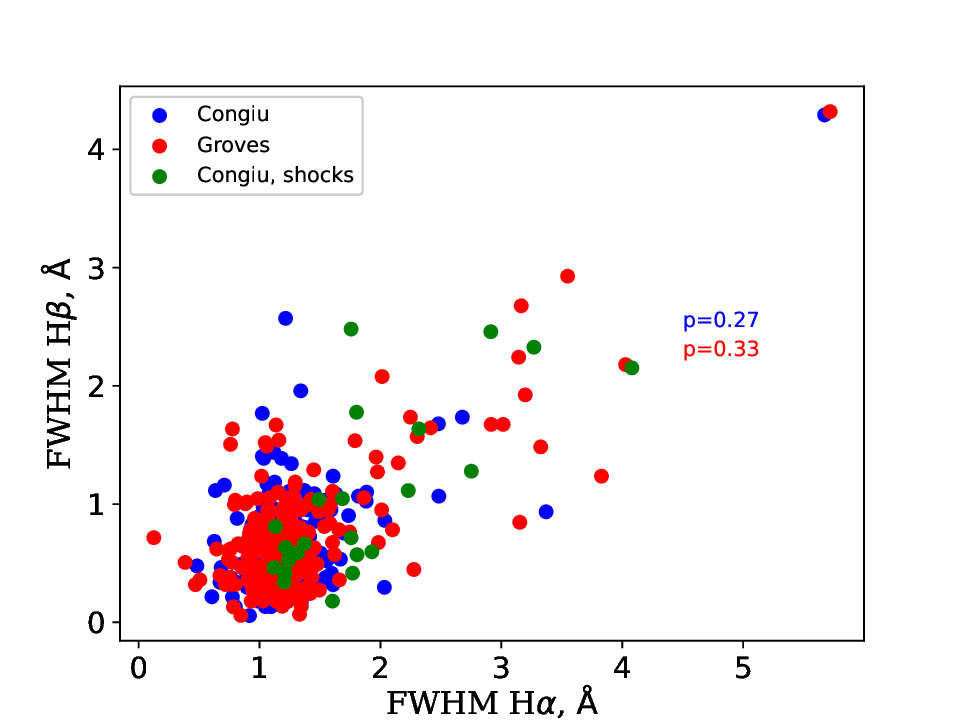}
\caption{\label{fig:HaHb} Relation between the FWHM of H$\beta$ lines on the H$\alpha$ lines for the selected regions: HII regions from the Groves catalog (red dots), HII regions from the Congiu catalog (blue dots), shock ionization regions from the Congiu catalog (green dots).}
\end{figure}


Figure~\ref{fig:HaHb} shows the relation between the FWHMs of H$\alpha$ and H$\beta$ lines for both catalogs. We see some excess of red dots around FWHM H$\alpha\approx1.5$\AA{} and H$\beta\approx1$\AA. This means that regions from the Groves catalog demonstrate higher values of the FWHM of H$\alpha$ and H$\beta$ compared to the Congiu catalog. It is noteworthy that if we take the regions of shock ionization from the Congiu catalog and plot them on the same figure (green dots), they fill in the same plot area. In other words, there are regions in the Congiu catalog with these values of H$\alpha$ and H$\beta$ FWHM, but they are classified as regions of shock ionization. This suggests that some of the regions from the Groves catalog can be misidentified as HII regions and are in fact shocks.

\begin{figure}[b]
\includegraphics[width=0.5\textwidth]{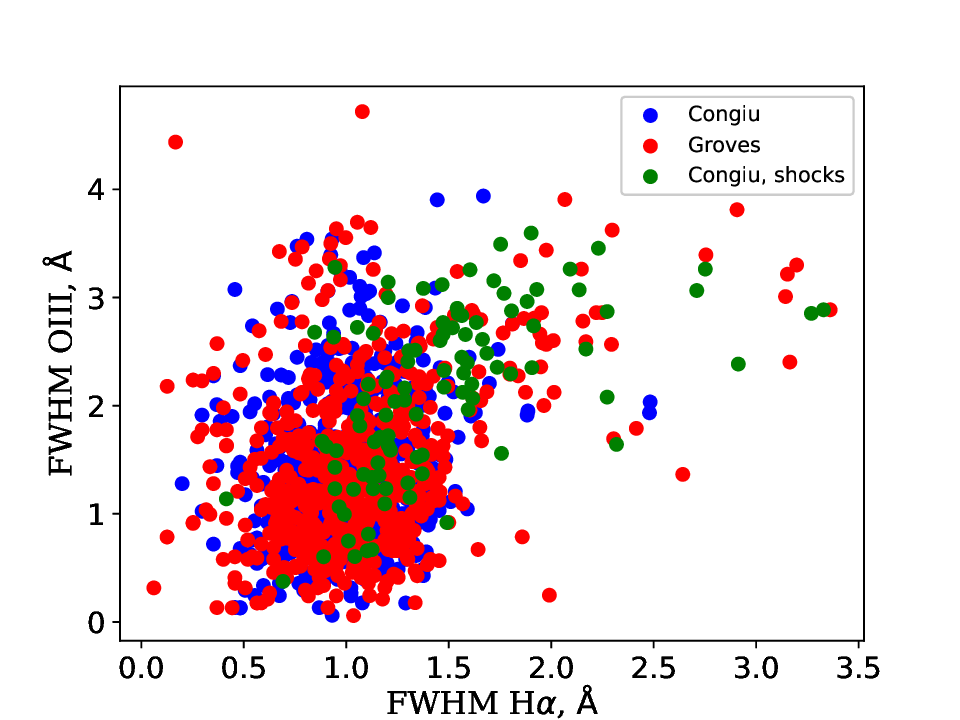}
\caption{\label{fig:HaOIII} Relation between the OIII and H$\alpha$ line FWHMs for star-forming regions of the Groves (red dots) and Congiu (blue dots) catalogs.}
\end{figure}

This is also seen in Fig.~\ref{fig:HaOIII}, where we relate FWHM values for H$\alpha$ and OIII lines. Again, there is some excess of red dots, corresponding to HII regions from the Groves catalog, having FWHM H$\alpha$ about 2\AA{} or greater, and this area is also populated by shocks from the Congiu catalog.

\begin{figure}[b]
\includegraphics[width=0.4\textwidth,angle =270]{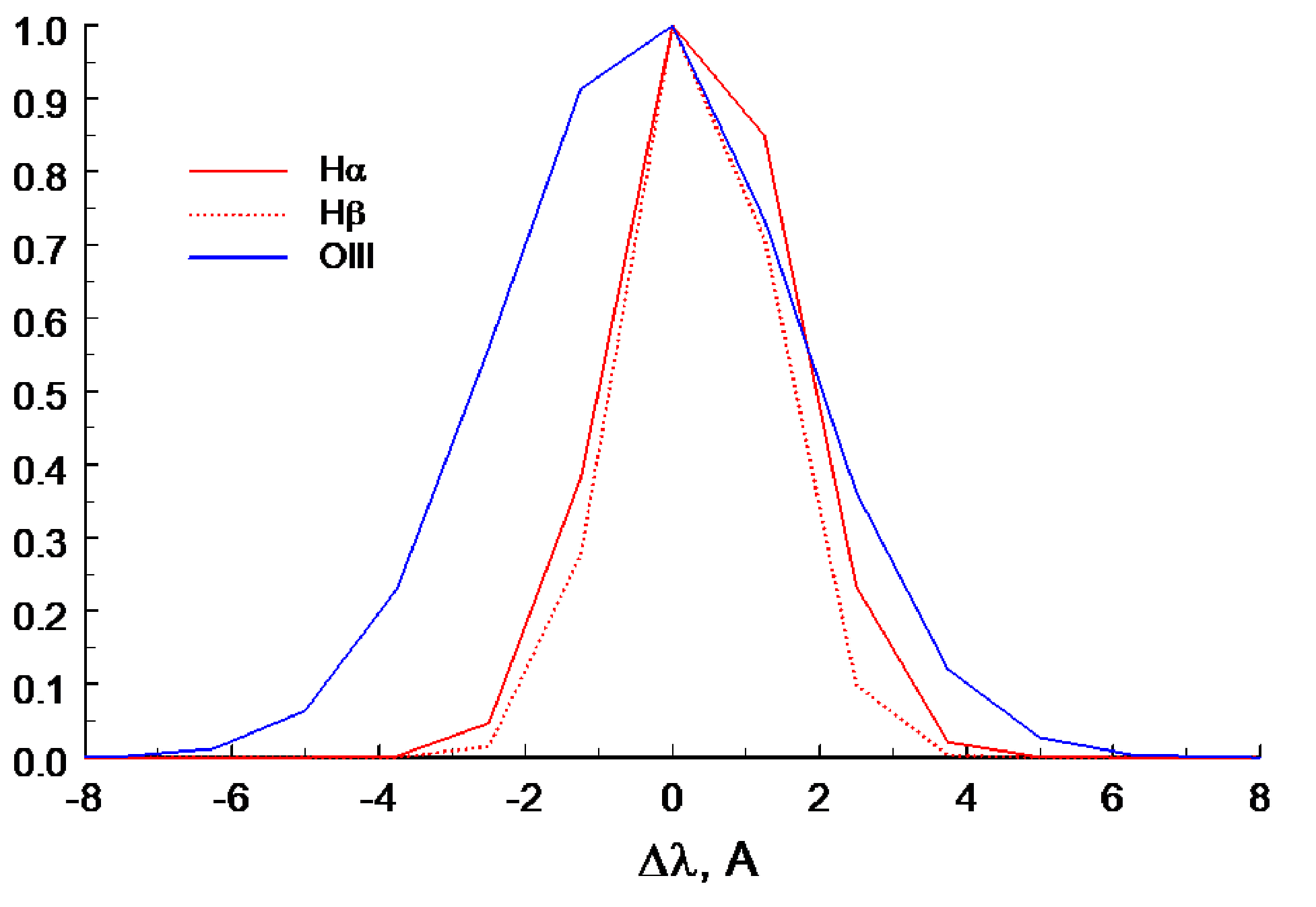}
\caption{\label{fig:spectra} Line profiles H$\alpha$, H$\beta$ and OIII of one of the regions of the galaxy NGC628. The line profiles were artificially brought to one center.}
\end{figure}

Fig.~\ref{fig:HaOIII} also shows that there are some regions in the Groves catalog with `typical' values of H$\alpha$ FWHM and higher than typical values of OIII FWHM ($>4$\AA). An example is shown in Fig.~\ref{fig:spectra}.  This indicates that in some cases breadths of these two lines can be quite different in the regions from this catalog, probably, originating in different conditions. Broad OIII lines may also indicate a significant contribution from impact excitation. Such regions are more likely to be classified as shocks rather than as classical HII regions \cite{14}. However, an extended statistical sample is required to finally confirm this conclusion.

It is interesting to note outstanding points in Fig.~\ref{fig:HaHb}. They correspond to the same region in both catalogs, which looks like a typical region in other respects. It deserves further study.

\section{\label{sec:level1}Conclusions}

When analyzing the two catalogs, we obtained the following conclusions:
\begin{enumerate}
\item Comparison of data from both catalogs shows good agreement between IR data and line data.
\item There are differences in the regions identified by the H$\alpha$ line data, spatial mismatches of the identified regions are observed.
\item A peculiar behavior of the OIII line data is observed, in different catalogs we see different behavior.
\item In the Groves catalog, some regions designated as HII regions may show signs of shock excitation
\end{enumerate}



\section*{FUNDING}
K. Smirnova has been supported by the Ministry of Science and Education of Russia, the FEUZ-2025-0003 project.

\section*{CONFLICT OF INTEREST}

The authors of this work declare that they have no conflicts of interest.



\begin{thebibliography}{}

\bibitem{1}
J. Roman-Duval et al., Astrophys. J. \textbf{797} 86 (2014).
https://doi.org/10.1088/0004-637X/797/2/86

\bibitem{2}
K. M. Sandstrom et al., Astrophys. J. \textbf{777}
5 (2013).
https://doi.org/10.1088/0004-637X/777/1/5

\bibitem{3}
R. C. Kennicutt et al., Publ. Astron. Soc. Pacif. \textbf{115}, 928 (2003).
https://doi.org/10.1086/376941

\bibitem{4}
A. K. Leroy et al., Astron. J. \textbf{137}, 4670 (2009). https://doi.org/10.1088/0004-6256/137/6/4670

\bibitem{5}
A. K. Leroy et al., Astrophys. J. Suppl. Ser. \textbf{257}, 43 (2021).
https://doi.org/10.3847/1538-4365/ac17f3

\bibitem{6}
A. K. Leroy et al., Astrophys. J. Suppl. Ser. \textbf{255}, 19 (2021).
https://doi.org/10.3847/1538-4365/abec80

\bibitem{7}
J. C. Lee et al., Astrophys. J. Lett. \textbf{944},  L17 (2023).
https://doi.org/10.3847/2041-8213/acaaae

\bibitem{8}
E. Emsellem et al.,  Astron. Astrophys. \textbf{659}, A191 (2022).
https://doi.org/10.1051/0004-6361/202141727

\bibitem{9}
K. Smirnova, M. Murga, D. Wiebe and A. Sobolev, ARep. \textbf{61}, 646 (2017). 
https://doi.org/10.1134/S1063772917070083. 

\bibitem{10}
K. Smirnova, D. Wiebe, ARep. \textbf{63}, 445 (2019). 
https://doi.org/10.1134/S1063772919060040.


\bibitem{11}
E. Congiu et al., Astronomy and Astrophysics \textbf{672}, A148 (2023). 
https://doi.org/10.1051/0004-6361/202245153

\bibitem{12}
B. Groves et al., Monthly Notices of the Royal Astronomical Society \textbf{520}, 4902 (2023).
https://doi.org/10.1093/mnras/stad114 

\bibitem{13}
J. P.Williams, E. J. de Geus \& L. Blitz, Astrophys. J. \textbf{428}, 693 (1994).
https://doi.org/10.1086/174279

\bibitem{14}
D. V. Oparin  \& A. V. Moiseev, Astrophysical Bulletin \textbf{73}, 298 (2018). 
https://doi.org/10.1134/S1990341318030045

\bibitem{15}
K. Kreckel et al., Astrophys. J. \textbf{887}, 80 (2019).
https://doi.org/10.3847/1538-4357/ab5115

\end{thebibliography}
\end{document}